# Fiber modal noise mitigation by a rotating double scrambler

G. Raskin*[a], D. Rogozin[ab], T. Mladenov[a], C. Schwab[b], D. Coutts[bc]
[a]KU Leuven, Institute of Astronomy, 3001 Leuven, Belgium; [b]Department of Physics and Astronomy, Macquarie University, NSW 2109, Australia; [c]MQ Photonics Research Centre, Macquarie University, NSW 2109, Australia

## ABSTRACT

Fiber modal noise is a performance limiting factor in high-precision radial velocity measurements with multi-mode fiber-fed high-resolution spectrographs. Traditionally, modal noise is mitigated by agitating the fiber, this way redistributing the light that propagates in the fiber over many different modes. However, in case of fibers with only a limited number of modes, e.g. at near-infrared wavelengths or in adaptive-optics assisted systems, this method becomes very inefficient. The strong agitation that would be needed stresses the fiber and can lead to focal ratio degradation. As an alternative approach, we propose to use a classic optical double scrambler and to rotate the scrambler's first fiber end during each exposure. Because of the rotating illumination pattern of the scrambler's second fiber, the modes that are excited vary continuously. This leads to very efficient averaging of the modal pattern at the fiber exit and to a strong reduction of modal noise. In this contribution, we present a prototype design and first laboratory results of the rotating double scrambler.

## 1. INTRODUCTION AND EXPERIMENT

Modal noise refers to the noise caused by the variable spatial distribution of monochromatic light at the output face of a multi-mode optical fiber. Changes in illumination, position, temperature, etc. of the fiber will produce a variable speckle pattern at the fiber output (Figure 2 a). When feeding a high-resolution spectrograph with a multi-mode fiber, modal noise effectively limits the radial velocity precision and the signal-to-noise ratio that can be achieved. Optical double scramblers [1] are routinely used to improve the illumination stability of high-resolution Doppler spectrographs. These scramblers consist of two identical lenses, connected to the fiber faces at a junction in the fiber train (Figure 1 a). The first lens images the first fiber face on the pupil of the second part and vice-versa. This way, the near and far field images of both fiber parts are swapped, and spatial and angular scrambling are greatly increased. Modal noise, unfortunately, is not reduced by the static optical double scrambler [2].

An efficient alternative to mechanic fiber agitation [3] is varying the input illumination of the fiber. To achieve this, we have equipped the first part of the double scrambler with a rotation mechanism. To avoid fiber wind-up, we continuously rotate between -180° and 180°. The second part of the scrambler, connected to the fiber leading to the spectrograph is fixed. For stability reasons, the rotating scrambler is mounted vertically (Figure 1 b). Both fiber ends and lenses are mounted on 5-axis adjustable mounts (XYZ/Tip/Tilt) for aligning their optical axes with the rotation axis of the mechanism. For our laboratory experiments, the scrambler is equipped with 2 achromats ($F = 6$ mm) and 60 μm circular fibers, illuminated by a 633 nm $F$/4 beam.

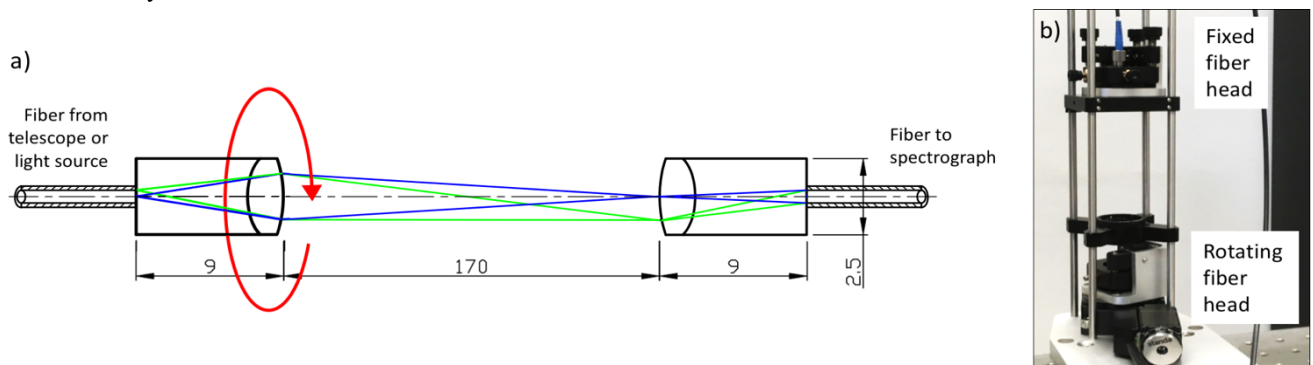

Figure 1. a) Layout and ray trace of the optical double scrambler; b) picture of the rotating scrambler setup.

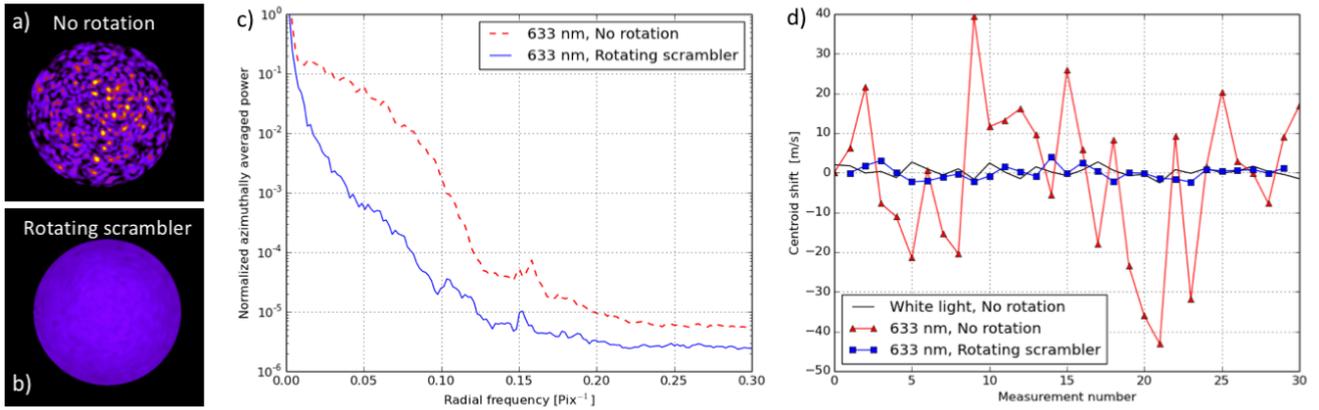

Figure 2. False color images of the fiber exit face a) without and b) with scrambler rotation (same color scale for both images); c) Azimuthally averaged power spectrum of images in a and b; d) Radial velocity scatter of a single spectral line with and without scrambler rotation (white light illumination is shown as noise floor reference).

## 2. RESULTS AND DISCUSSION

Microscope images of the fiber exit face are shown in Figure 2 a & b. Without scrambler rotation, the modal speckle pattern is very pronounced while it is almost completely averaged out during a 1-minute integration over five 360° rotations of the scrambler. In Figure 2 c, we plot the azimuthally averaged power spectral density of these fiber face images. At spatial frequencies corresponding with typical modal speckle size, the rotating scrambler reduces the structure in the image by about two orders or magnitude.

Changes in the speckle pattern also affect the centroid position of the fiber image, resulting in radial velocity errors. We took a series of measurements, for each data point modifying the position of the input fiber in order to change the illumination pattern. For white light, this results in a centroid scatter of 1.4 m/s, corresponding with the measurement accuracy of our setup. Illuminated with monochromatic light, this increases to 1.6 m/s and 18 m/s, respectively with and without rotation of the scrambler. These scatter values correspond with the centroid error for a single spectral line at a resolution of $R = 100\,000$. The radial velocity error of a complete spectrum, containing thousands of lines, would be reduced to a level below 5 cm/s.

Alignment of the rotating optical double scrambler is a tricky aspect of this setup. Both the images and the pupils of both fiber faces need to stay precisely aligned during rotation. We measured a peak-to-peak flux variation of 5% over a complete revolution, indicating there is still room for a small improvement. Clearly, the mechanical design of the rotation mechanism is critical as the alignment needs to be maintained over a long period and over many rotation cycles. As a more elegant alternative to rotating one fiber end back and forth, we could also install a dove prism on a rotation stage between both fixed fiber ends. This would allow smoother operation and a constant rotation speed at the cost of a small throughput loss.

This solution for modal noise mitigation in multi-mode fibers can be generally applied to any high-resolution spectrograph that is equipped with an optical double scrambler. It is efficient for both stellar spectra as well as for highly coherent wavelength calibration sources like the laser frequency comb. It is especially attractive for near-infrared or AO-assisted systems where the number of fiber modes is limited and mechanical fiber agitation is insufficient for obtaining extreme-precision radial velocity measurements.

*gert.raskin@kuleuven.be